\documentclass[%
 aip,
 amsmath,amssymb,
 reprint,%
]{revtex4-1}

\usepackage{graphicx}
\usepackage{dcolumn}
\usepackage{bm}

\usepackage[utf8]{inputenc}
\usepackage[T1]{fontenc}
\usepackage{mathptmx}
\usepackage{etoolbox}
\usepackage{gensymb}
\usepackage{bigints}
\usepackage{tablefootnote}

\makeatletter
\def\@email#1#2{%
 \endgroup
 \patchcmd{\titleblock@produce}
  {\frontmatter@RRAPformat}
  {\frontmatter@RRAPformat{\produce@RRAP{*#1\href{mailto:#2}{#2}}}\frontmatter@RRAPformat}
  {}{}
}%
\makeatother
\begin{document}

\preprint{AIP/123-QED}

\title{Dielectric Reliability and Interface Trap Characterization in MOCVD grown In-situ Al$_2$O$_3$ on $\beta$-Ga$_2$O$_3$}

\author{Saurav Roy$^*$}
\email{sauravroy@ucsb.edu, sriramkrishnamoorthy@ucsb.edu}

\affiliation{%
Materials Department, University of California Santa Barbara, Santa Barbara, CA, 93106 
}%

\author{Arkka Bhattacharyya}%

\affiliation{%
Materials Department, University of California Santa Barbara, Santa Barbara, CA, 93106 
}%

\author{Carl Peterson}
\affiliation{%
Materials Department, University of California Santa Barbara, Santa Barbara, CA, 93106 
}%

\author{Sriram Krishnamoorthy$^*$}
\affiliation{%
Materials Department, University of California Santa Barbara, Santa Barbara, CA, 93106 
}%

\date{\today}

\begin{abstract}
In this article, we investigate the in-situ growth of Al$_2$O$_3$ on $\beta$-Ga$_2$O$_3$ using metal-organic chemical vapor deposition (MOCVD) at a high temperature of 800°C. The Al$_2$O$_3$ is grown within the same reactor as the $\beta$-Ga$_2$O$_3$, employing trimethylaluminum (TMAl) and O$_2$ as precursors without breaking the vacuum. We characterize the shallow and deep-level traps through stressed capacitance-voltage (C-V) and photo-assisted C-V methods. The high-temperature deposited dielectric demonstrates an impressive breakdown field of approximately 10 MV/cm. Furthermore, we evaluate the reliability and lifetime of the dielectrics using time-dependent dielectric breakdown (TDDB) measurements. By modifying the dielectric deposition process to include a high-temperature (800°C) thin interfacial layer and a low-temperature (600°C) bulk layer, we report a 10-year lifetime under a stress field of 3.5 MV/cm along a breakdown field of 7.8 MV/cm.
\end{abstract}

\maketitle

The future advancement of $\beta$-Ga$_2$O$_3$-based high-power electronics \cite{green2022beta, pearton2018review} hinges on the development of suitable dielectrics and the refinement of deposition techniques to satisfy stringent gate insulation \cite{peterson2024kilovolt, li2019single} and device passivation requirements \cite{bhattacharyya20224, bhattacharyya2021multi, saha2023scaled, saha2022temperature}. Typically, $\beta$-Ga$_2$O$_3$ materials are grown in a growth reactor \cite{murakami2014homoepitaxial, bhattacharyya2020low, mauze2020sn} and then transferred to another system for dielectric deposition \cite{masten2019ternary, jian2020deep}. During this transfer, contaminants from ambient air can accumulate on the surface. Removing these contaminants requires extensive chemical pre-treatments \cite{9870683}, which lengthen the processing time and may introduce undesired side effects that compromise interface quality and also may not be compatible with prior process steps.

These issues can be mitigated by using an in-situ growth approach, where the dielectric is deposited in the same equipment as the $\beta$-Ga$_2$O$_3$ layers. This method has been successfully applied to the in-situ growth of low-temperature dielectrics on III-nitrides \cite{liu2014metalorganic} and $\beta$-Ga$_2$O$_3$ using metalorganic chemical vapor deposition (MOCVD) \cite{roy2021situ, jian2021characterization} and molecular beam epitaxy (MBE) \cite{dhara2023plasma}.

\begin{figure}[t]
\centering
\includegraphics[width=3.5in,height=15cm, keepaspectratio]{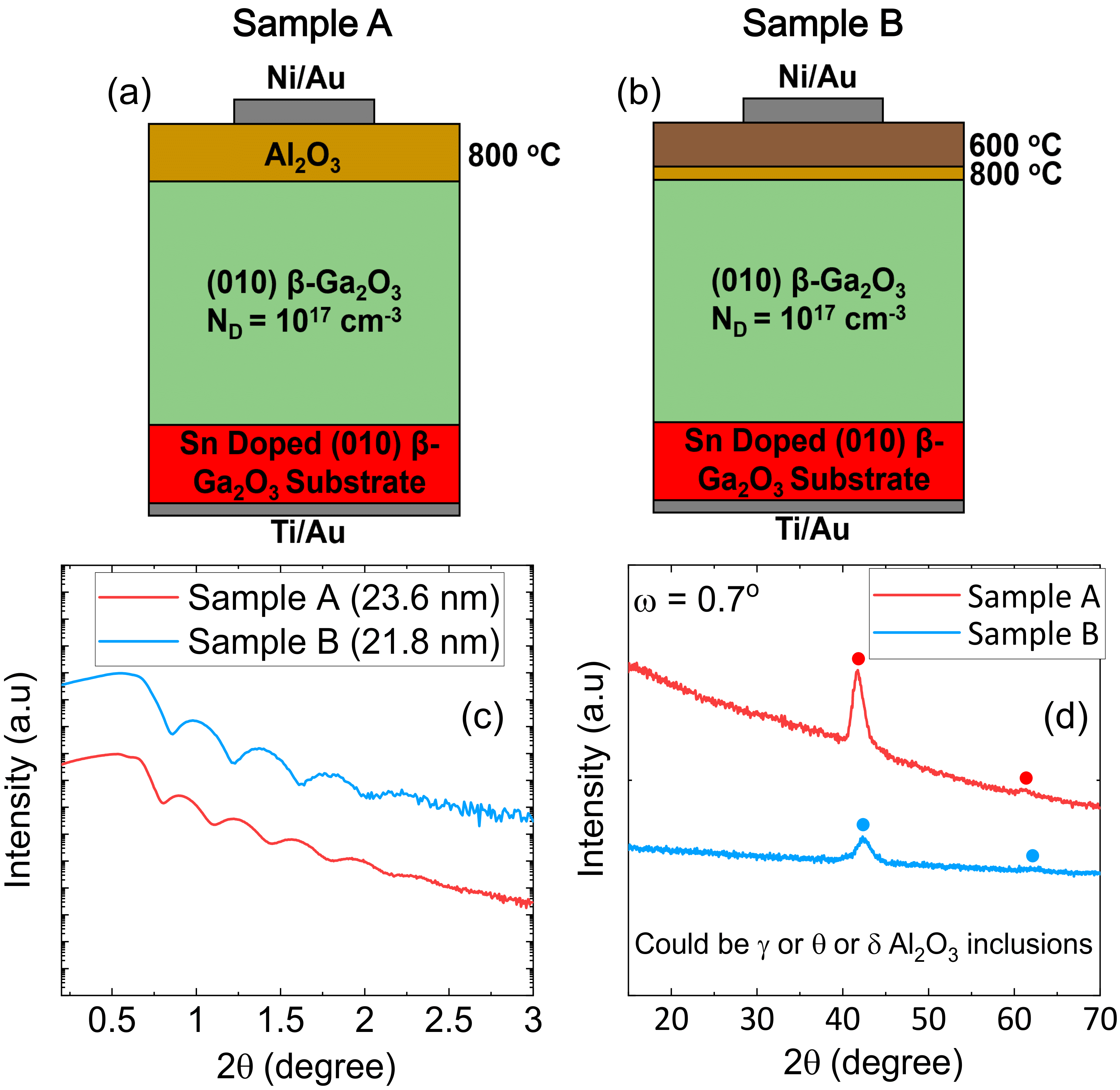}
\caption{Schematic diagram of the fabricated MOSCAP structures with (a) Sample A (800 $\degree$C) and (b) Sample B (800/600 $\degree$C). (c) X-ray reflectivity plot and (d) GIXRD spectra for Sample A and Sample B.}
\label{fig1}
\vspace{-0.5cm}
\end{figure}

In this letter, we report on the in-situ MOCVD growth and capacitance-voltage (C-V) characterization of Al$_2$O$_3$ on (010) oriented $\beta$-Ga$_2$O$_3$ metal-oxide-semiconductor capacitors (MOSCAPs). Two different dielectrics are grown, one at 800 $\degree$C (Fig. \ref{fig1} (a)) and another with 800 $\degree$C interfacial layer and the bulk grown at 600 $\degree$C (Hybrid) (Fig. \ref{fig1} (b)). Additionally, we present findings on the quality, reliability, and lifetime analysis of in-situ MOCVD-grown Al$_2$O$_3$ dielectrics grown at various temperatures. This includes evaluations of interface-state density, leakage current, breakdown strength, and time-dependent dielectric breakdown.

The growth of a 500 nm $\beta$-Ga$_2$O$_3$ epilayer on Sn-doped bulk (010) $\beta$-Ga$_2$O$_3$ substrates (NCT Japan) was conducted using an Agnitron MOVPE reactor with a far injection showerhead design \cite{ranga2020delta}. Prior to loading the sample into the growth chamber, the substrates were cleaned with acetone, methanol, and deionized (DI) water, followed by a 15-minute dip in hydrofluoric acid (HF). Triethylgallium (TEGa) and O$_2$ were used as precursors, argon (Ar) as the carrier gas, and diluted silane as the dopant. The growth process was performed at a low temperature of 600 $\degree$C and a chamber pressure of 60 Torr. The gas flow rates for TEGa and O$_2$ were set at 65 sccm and 800 sccm, respectively, with an Ar flow rate of 1100 sccm. A silane flow rate of 0.06 nmol/min was used to achieve a silicon doping concentration of 10$^{17}$ cm$^{-3}$. Under these conditions, a $\beta$-Ga$_2$O$_3$ growth rate of 6.3 nm/min was achieved.

After the $\beta$-Ga$_2$O$_3$ growth, a 90-second purge step was implemented to remove all unreacted precursors from the reactor. During this step, the reactor pressure was reduced from 60 to 15 Torr while the substrate temperature was increased from 600 $\degree$C to 800 $\degree$C. Additionally, the O$_2$ flow rate was decreased from 800 to 500 sccm. Following this purge step, an in-situ Al$_2$O$_3$ layer was grown at a rate of 1.2 nm/min using the trimethylaluminum (TMAl) precursor at a flow rate of 5.33 $\mu$mol/min and O$_2$ at a flow rate of 500 sccm. The growth rate of the Al$_2$O$_3$ layer was determined to be 4.7 nm/min. This is referred to as sample A.

To grow the hybrid dielectric (800 $\degree$C thin interfacial layer/600 $\degree$C bulk), a 5 nm thin interfacial layer was first grown at 800 $\degree$C. After this, the growth temperature was ramped down from 800 $\degree$C to 600 $\degree$C over the course of 5 minutes. Following this temperature ramp, the bulk of the dielectric was grown at 600 $\degree$C, using the same gas flow rates and chamber conditions as for the 800 $\degree$C dielectric. The growth rate of the dielectric at 600 $\degree$C was determined to be 1.2 nm/min. This is referred to as sample B. The thicknesses of the dielectric layers were measured using X-ray reflectivity (XRR) as shown in Fig. \ref{fig1} (c), resulting in values of 23.6 nm for the 800°C dielectric and 21.8 nm for the hybrid dielectric.

\begin{figure}[t]
\centering
\includegraphics[width=3.5in,height=16cm, keepaspectratio]{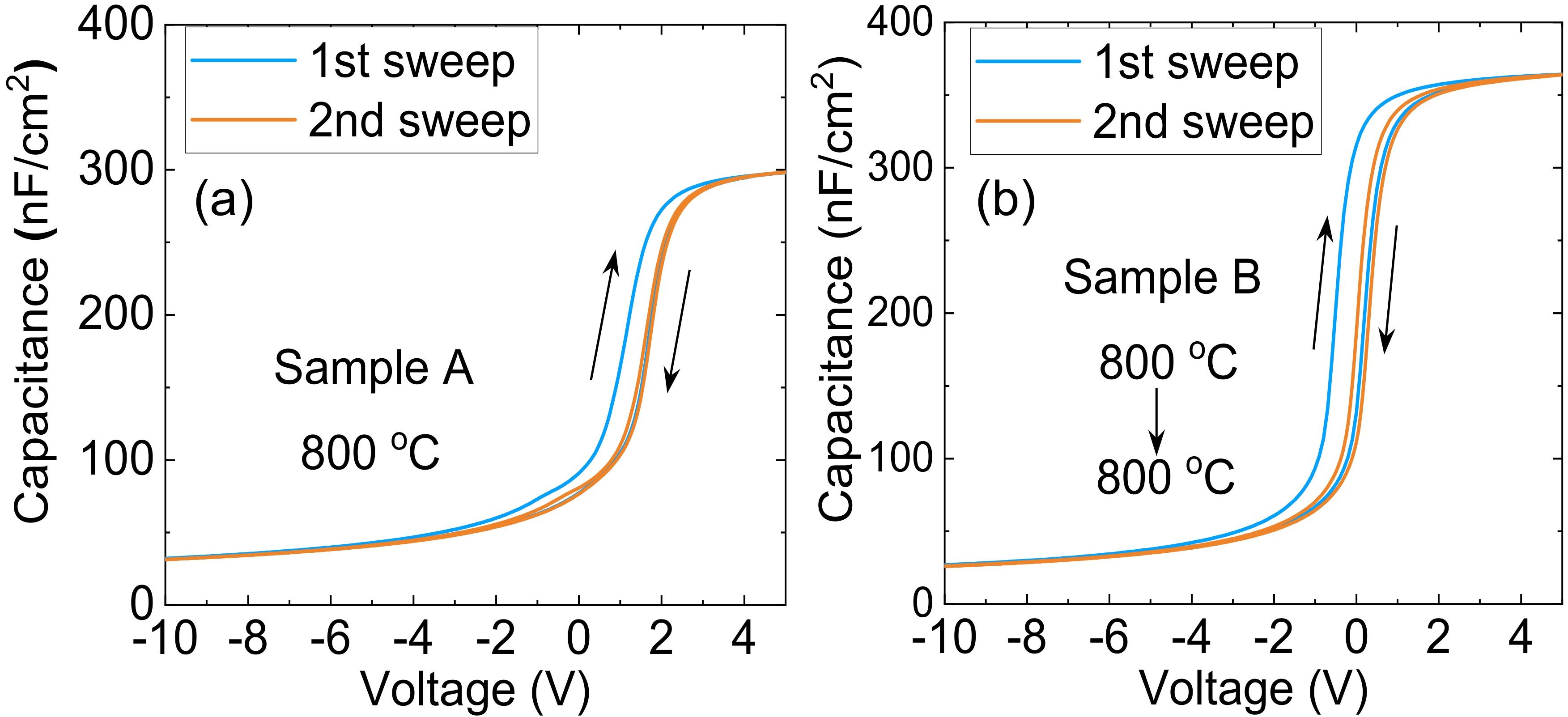}
\caption{First and second sweep of capacitance vs voltage characteristics for (a) Sample A and (b) Sample B. In each case, during the first D-A sweep, MOSCAPs were subjected to a 10 minutes of accumulation stress.}
\label{fig3}
\vspace{-0.5cm}
\end{figure}

Grazing incident X-ray measurements (GIXRD) were conducted on the dielectrics grown using metal-organic chemical vapor deposition (MOCVD) to analyze the crystallinity of the MOCVD grown Al$_2$O$_3$ layers. The measurements were performed at a fixed incident angle of $0.7\degree$, as depicted in Fig. \ref{fig1}(d). The resulting GIXRD spectra were then compared against the International Centre for Diffraction Data (ICDD) Powder Diffraction File (PDF) database to determine the phase of Al$_2$O$_3$. Dielectrics in sample A and B exhibited some degree of crystallinity, possibly in the $\theta$ or $\gamma$ phase. However, the presence of the $\delta$ phase cannot be entirely ruled out as the major diffraction peaks of the $\delta$ phase closely resemble those of the $\gamma$ or $\theta$ phase. Further microstructure analysis using transmission electron microscopy will be reported elsewhere.

After the growth of the Al$_2$O$_3$ layer, Ti (50 nm)/Au (100 nm) ohmic contacts were sputter-deposited on the backside of the MOSCAPs. Finally, circular Ni (10 nm)/Au (50 nm) gate contacts were patterned using standard photolithography and deposited on the Al$_2$O$_3$ surface via E-beam evaporation.

The detailed C-V measurement techniques and the extraction of trap densities are described elsewhere \cite{roy2021situ, liu2013fixed}. Using a direct-current (DC) bias range of (-10 V, 5 V) and a sweep rate of 200 mV/s, a MOSCAP was initially swept from depletion to accumulation (D to A) and held in accumulation for 10 minutes before being swept back to depletion (A to D). It was then swept from D to A and immediately back from A to D for a second time. The alternating-current (AC) measurement bias and frequency were set at 30 mV and 1 MHz, respectively. Fig. \ref{fig3}(a) and \ref{fig3}(b) present the first and second pairs of D to A and A to D sweeps measured on MOSCAPs with sample A and sample B, respectively.

\begin{figure}[t] 
\centering
\includegraphics[width=3.5in,height=15cm, keepaspectratio]{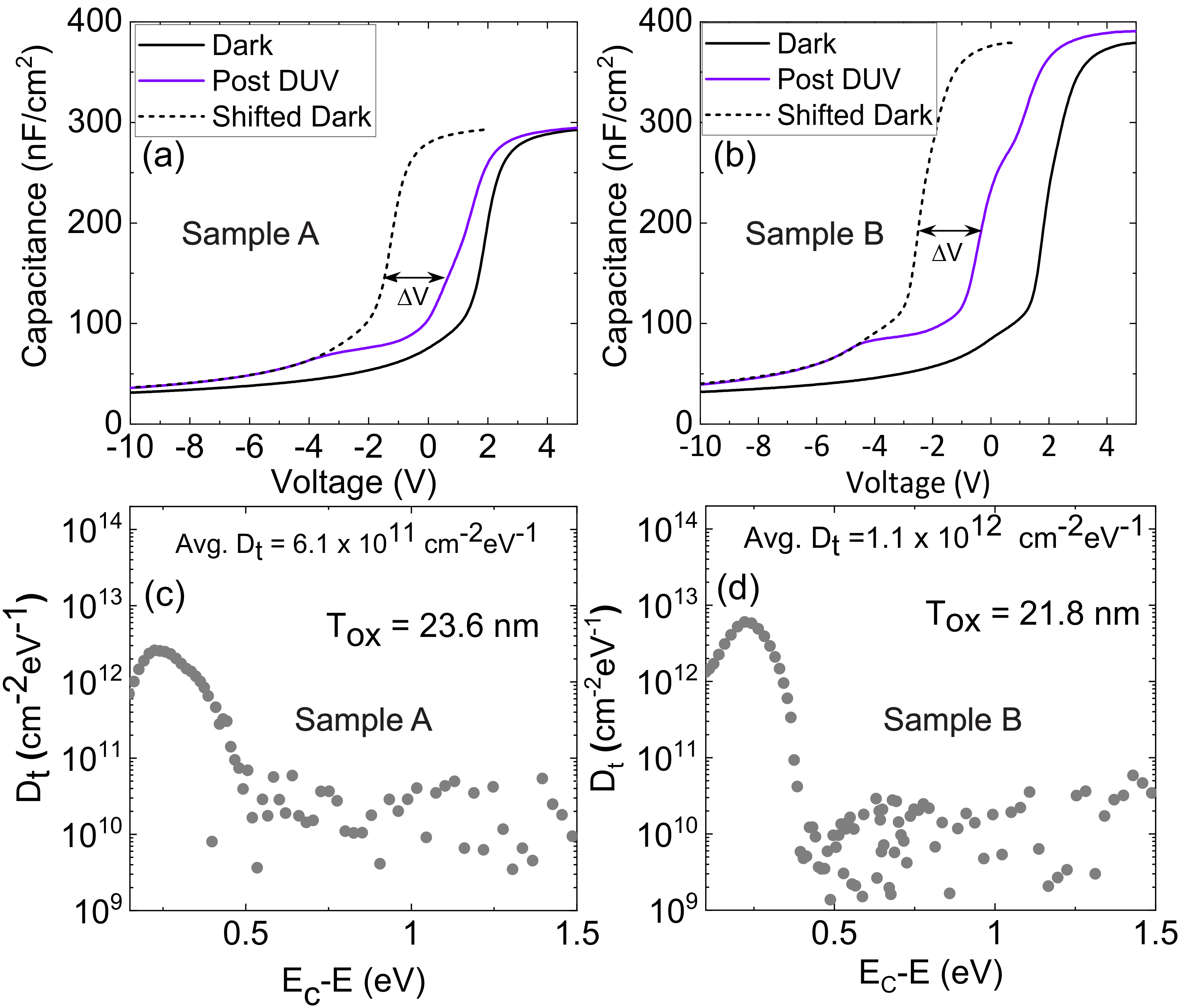}
\caption{UV assisted CV characteristics and Trap density vs energy distribution plots respectively of the MOSCAPs on (a) (c) Sample A and (b) (d) Sample B.}
\label{fig4}
\vspace{-0.5cm}
\end{figure}

\begin{table*} [t]
\caption{Near-interface trap densities of the MOSCAPs}
\vspace{-0.2cm}
\begin{center}
\begin{minipage}{14cm}
\begin{tabular}{|c|c|c|c|c|c|c|}
\hline
 Dielectric Type & $\epsilon_{ox}$ & $\Delta Q_{T1}$ ($cm^{-2}$)\footnote{$\Delta Q_{T1}$ = Amount of fast and initially empty slow near-interface states} & $\Delta Q_{T2}$ ($cm^{-2}$) \footnote{$\Delta Q_{T2}$ = Amount of fast near-interface states}  & $\Delta Q_{T1}-\Delta Q_{T2}$ ($cm^{-2}$) \footnote{$\Delta Q_{T1}-\Delta Q_{T2}$ = Amount of initially empty slow near-interface states}  & Avg. $D_t$ ($cm^{-2}eV^{-1}$) \footnote{Avg. $D_t$ extracted from deep UV CV measurements.}\\ 
 \hline \hline
 \cite{roy2021situ} 600 $\degree C$  & 8.3 & $-1.7\times 10^{12}$ & $-1.2\times 10^{12}$ & $-5\times 10^{11}$ & $3.2\times 10^{12}$ \\  
 \hline
 Sample A (800 $\degree C$) & 8.32 & $-8.3\times 10^{11}$ & $-1.5\times 10^{11}$ & $-6.8\times 10^{11}$ & $6.1\times 10^{11}$ \\  
 \hline
 Sample B (800/600 $\degree C$) & 9.35 & $-1.64\times 10^{12}$ & $-2.74\times 10^{11}$ & $-1.36\times 10^{11}$ & $1.02\times 10^{12}$ \\ 
 \hline
 \hline
\end{tabular}
\end{minipage}
\label{t1}
\end{center}
\end{table*}

In this study, we focus on interface traps and border traps, which can be identified by the hysteresis they induce in the C-V curve. However, distinguishing between interface states and border traps using only C-V measurements is not straightforward, so we treat them collectively as near-interface states. The hysteresis ($\Delta V_{FB}$) is quantified as the difference in flat band voltage between each pair of depletion-to-accumulation (D to A, $V_{FB}^{D-A}$) and accumulation-to-depletion (A to D, $V_{FB}^{A-D}$) sweeps as shown in Fig. \ref{fig3}(a) and (b). After the first D to A sweep, the MOSCAPs are held in accumulation for 10 minutes, allowing electrons to fill all the empty near-interface states and remain trapped. During any D to A sweep, electrons can also fill the fast near-interface states, but these electrons are not released at the same rate during the subsequent A to D sweep. Consequently, the first hysteresis ($\Delta V_{FB1}$) is influenced by both fast and initially empty slow near-interface states, while the second hysteresis ($\Delta V_{FB2}$) is influenced only by the fast near-interface states \cite{liu2014situ}. Without the presence of any additional charge components, both $\Delta V_{FB1}$ and $\Delta V_{FB2}$ should be positive due to electron trapping, with $\Delta V_{FB1}$ being larger than $\Delta V_{FB2}$. The trapped electron density ($\Delta Q_{T}$) associated with $\Delta V_{FB}$ can be estimated from,

\begin{equation}
\label{eq1}
   \Delta Q_T = -\frac{C_{ox}\Delta V_{FB}}{q}
\end{equation}

Using (\ref{eq1}), we can determine the amount of fast and initially empty slow near-interface states ($\Delta Q_{T1}$) as well as the fast interface states ($\Delta Q_{T2}$). Also, the difference, $\Delta Q_{T1} - \Delta Q_{T2}$, represents the densities of the fast and initially empty slow near-interface states. From these measurements, trap densities and dielectric constants are extracted and summarized in Table \ref{t1}. For comparison, the trap densities of the 600 $\degree$C dielectrics from our previous work \cite{roy2021situ} are also included in Table \ref{t1}.

The MOSCAP with the sample A exhibits fewer fast and initially empty slow near interface states ($\Delta Q_{T1}$=$-8.3\times 10^{11}$ $cm^{-2}$) and also the lower amount of fast interface states ($\Delta Q_{T2}$=$-1.5\times 10^{11}$ $cm^{-2}$) compared to the 600 $\degree$C dielectric as can be seen from Table \ref{t1}. This improvement is likely due to the enhanced interface crystallization at 800 $\degree$C, which reduces the number of Al dangling bonds at the dielectric/semiconductor interface. However, dielectric growth at higher temperatures can also lead to bulk nanocrystalline inclusions, raising long-term reliability concerns. To mitigate these issues, a hybrid dielectric structure with a 5nm interfacial layer grown at 800 $\degree$C (5nm) and bulk dielectric grown at 600 $\degree$C (sample B) is more favorable. $\Delta Q_{T1}$ and $\Delta Q_{T2}$ of the sample B is also found to be lower than 600 $\degree$C dielectric as shown in Table \ref{t1}. Higher values of $\Delta Q_{T1}$ and $\Delta Q_{T2}$ for the sample B compared to the sample A is likely due to the near interface states at the 800/600 $\degree$C dielectric interface.

\begin{figure}[t]
\centering
\includegraphics[width=3.5in,height=15cm, keepaspectratio]{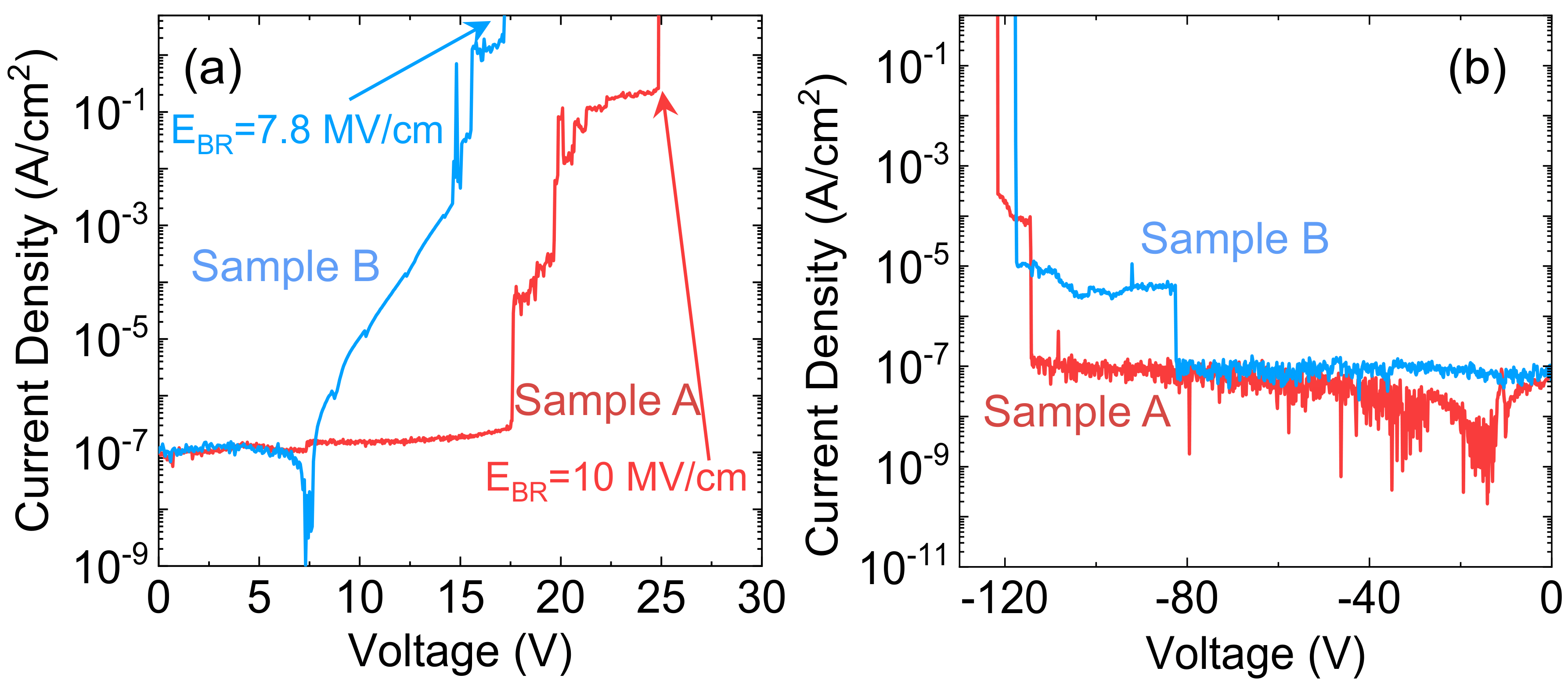}
\caption{(a) Forward and (b) reverse current-voltage characteristics for MOSCAPs on Sample A and Sample B.}
\label{fig5}
\vspace{-0.5cm}
\end{figure}

However, the density of slow and initially filled interface states cannot be extracted using the above-mentioned procedure due to the long emission times of the trap states. Therefore, an ultra-violet (UV) assisted CV method \cite{swenson2009photoassisted, yeluri2013capacitance} is employed to determine the density of these trap states and their energy dependence. The detailed procedure is described in \cite{roy2021situ} and is briefly summarized here. The MOSCAPs were initially taken from depletion to accumulation and kept in accumulation for 10 minutes to fill all the interface states. After this period, the MOSCAPs were swept back to depletion without any optical excitation. This "dark" curve is considered the ideal CV curve, assuming all interface states are filled with electrons and are neutral. Next, the MOSCAPs were held at depletion and exposed to UV light using a deep ultraviolet (DUV) light-emitting diode (LED) with a wavelength of 254 nm for 5 minutes. After the UV light was turned off, the MOSCAPs remained in depletion for an additional 15 minutes, followed by another sweep from depletion to accumulation. The dark curve was then horizontally shifted to align with the post-DUV CV curve, creating the reference CV curve. The total density of all the interface states was calculated from the difference between the post-DUV and reference CV curves ($\Delta V$). The energy dependence of these states was determined by the amount of band bending (dielectric/semiconductor surface potential) that occurred during the bias sweep. The deep UV-assisted CV plots for sample A and sample B are shown in Fig. \ref{fig4}(a-b).

\begin{figure}[t]
\centering
\includegraphics[width=3.5in,height=15cm, keepaspectratio]{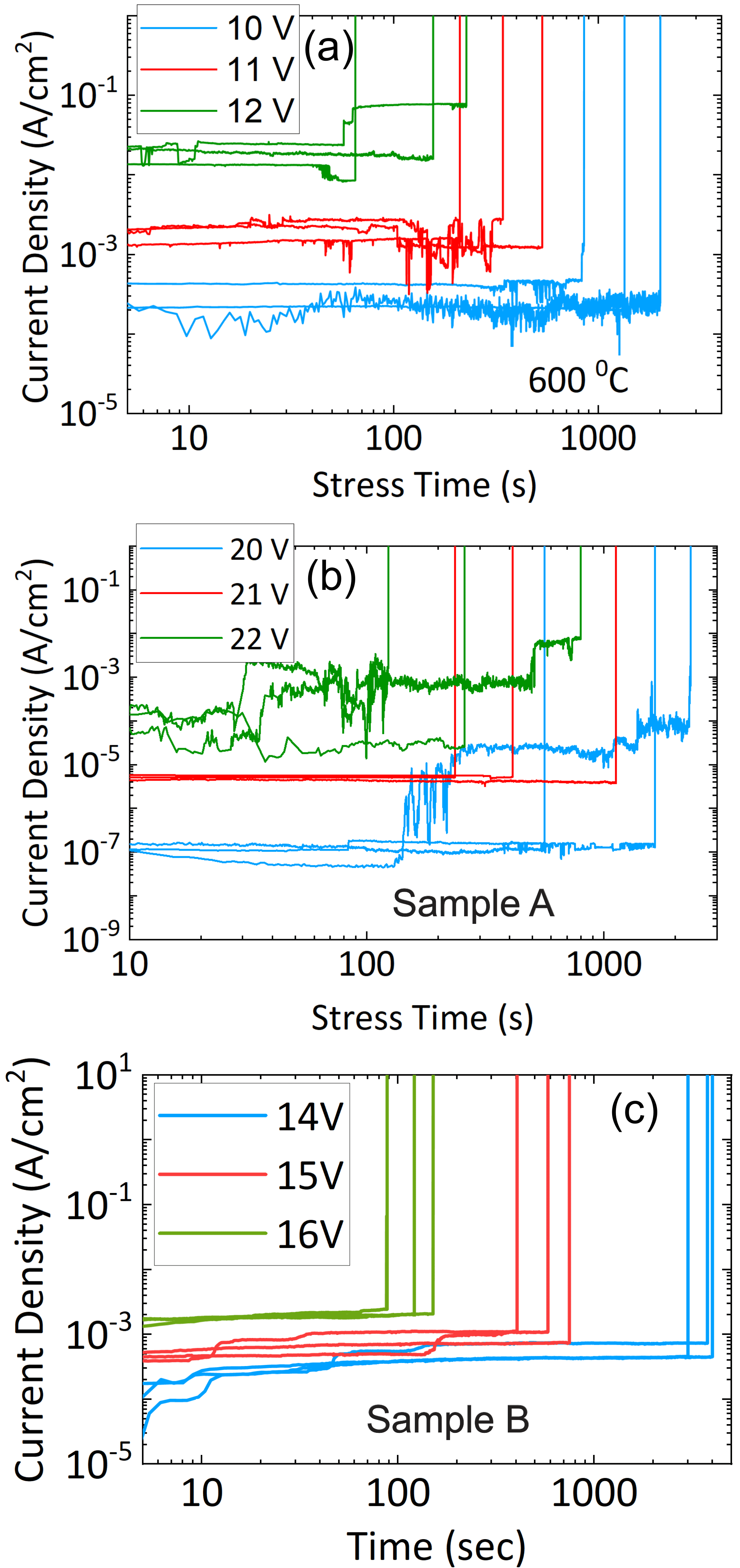}
\caption{Constant-voltage-stress current density vs time plot performed at three different voltages on the MOSCAPs on (a) 600 $\degree$C \cite{roy2021situ} (b) Sample A, and (c) Sample B.}
\label{fig6}
\vspace{-0.5cm}
\end{figure}

\begin{figure}[t]
\centering
\includegraphics[width=3.5in,height=15cm, keepaspectratio]{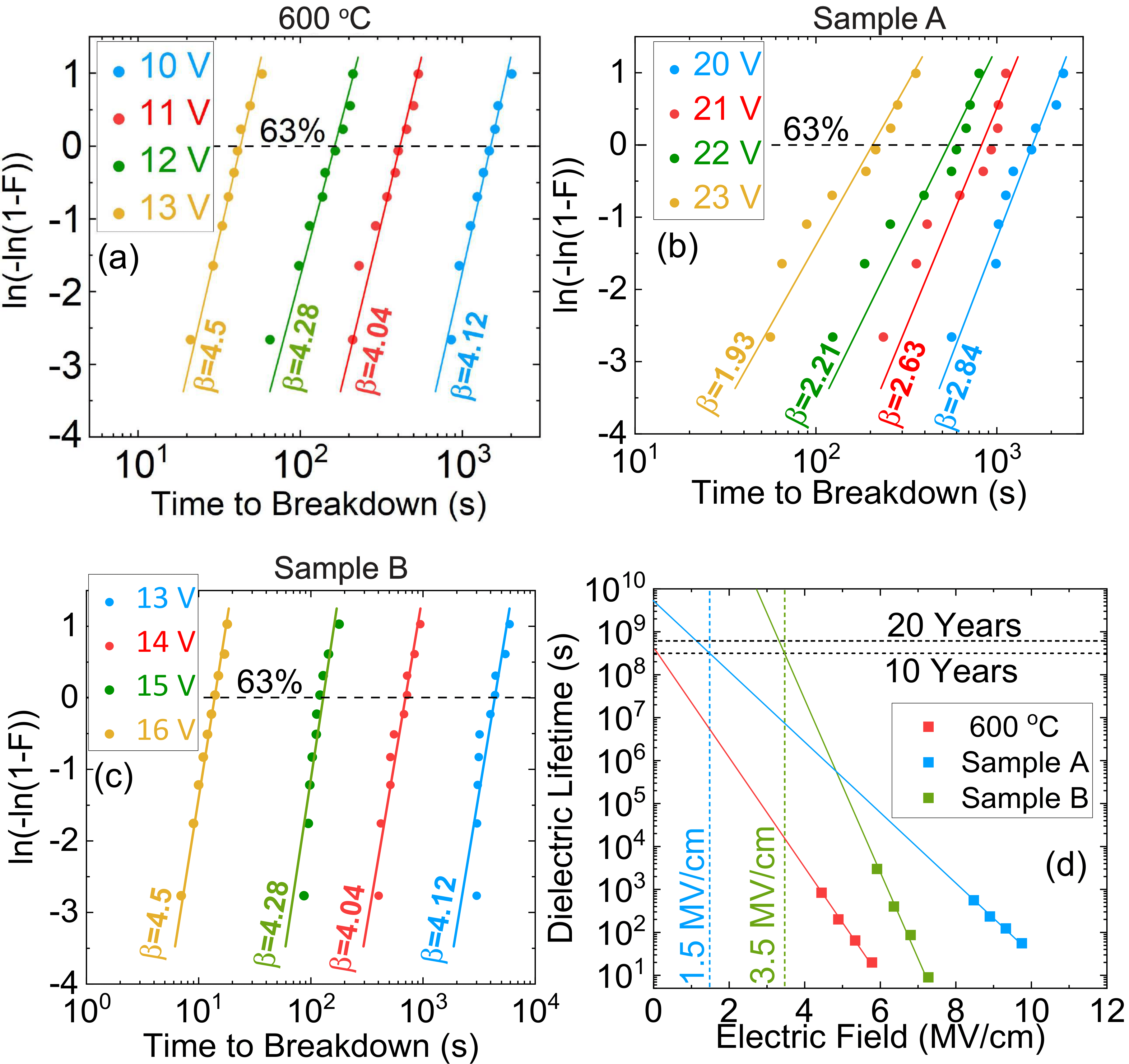}
\caption{Weibull distributions of measured lifetimes at room temperature with different voltage stress for (a) 600 $\degree$C \cite{roy2021situ} (b) Sample A, and (c) Sample B. (d) Dielectric lifetime vs oxide electric field for the three different dielectrics}
\label{fig7}
\vspace{-0.5cm}
\end{figure}

Fig. \ref{fig4}(c) to (d) depicts the relationship between trap density (D$_t$) and the trap energy distribution for the MOSCAPs with sample A and sample B. The presence of a positive valence band offset at the Al$_2$O$_3$/$\beta$-Ga$_2$O$_3$ results in the accumulation of additional holes at the Al$_2$O$_3$/$\beta$-Ga$_2$O$_3$ interface, which do not recombine with the interface traps. Notably, the increase in D$_t$ near the conduction band edge can be attributed to these accumulated holes. To obtain the average D$_t$ for the MOSCAPs, we exclude the contribution from holes by disregarding the rise in D$_t$. The average D$_t$ values for the different MOSCAPs are listed in Table \ref{t1}. The MOSCAP with sample A exhibits the lowest average $D_t$ density ($6.1 \times 10^{11}$ cm$^{-2}$eV$^{-1}$) among the three MOSCAPs. Additionally, the MOSCAP with sample B also shows a lower average $D_t$ ($1.02 \times 10^{12}$ cm$^{-2}$eV$^{-1}$) compared to the MOSCAP grown at 600$\degree$C dielectric ($3.2 \times 10^{12}$ cm$^{-2}$eV$^{-1}$). Notably, the $D_t$ values for both sample A and sample B are lower than the $\Delta Q_{T1}$ and $\Delta Q_{T2}$ values, which likely indicates a different origin for the fast and initially empty slow interface states.

To determine the breakdown voltages of the MOSCAPs under forward and reverse bias, current-voltage (IV) measurements were performed, as shown in Fig. \ref{fig5}. Sample A exhibits a breakdown voltage of 25V, corresponding to an impressive breakdown field of approximately 10 MV/cm under forward bias, as illustrated in Fig. \ref{fig5}(a). For sample B, the breakdown voltage is around 17V, translating to a breakdown field of about 7.8 MV/cm.

The reverse breakdown voltage, derived from the IV characteristics in Fig. \ref{fig5}(b), is approximately 120V for MOSCAPs with both sample A and sample B. In the accumulation case, with a positive potential applied to the dielectric, the entire voltage drop occurs across the Al$_2$O$_3$ layer, thus the breakdown can be attributed to the dielectric. However, under reverse bias, the total applied voltage is distributed across both the Al$_2$O$_3$ and Ga$_2$O$_3$ regions. Consequently, a significant electric field is present in the Ga$_2$O$_3$, indicating that the reverse breakdown cannot be solely attributed to the dielectric breakdown. Hence, the forward breakdown is a good indication of the dielectric material quality.

To characterize the reliability of the in-situ dielectrics, time-dependent dielectric breakdown tests \cite{hiraiwa2018time, bisi2016quality} were performed under constant voltage stress. The currents in the 600$\degree$C, 800$\degree$C (sample A), and hybrid dielectric (sample B) MOSCAPs under constant stressing voltages at room temperature are displayed in Fig. \ref{fig6}(a)-(c). For each dielectric sample, the MOSCAPs were stressed at three different voltages close to the breakdown voltage of the devices, with 10 devices for each sample stressed at the same voltage to obtain sufficiently good statistics. The 600$\degree$C and sample B exhibit hard breakdown during stress, indicated by a sudden rise in current marking the breakdown. However, sample A exhibits some pre-breakdown current increases in many devices, indicating soft breakdown, which likely results from leakage through grain boundaries of polycrystalline regions. The time-to-breakdown for dielectrics can be well approximated using a Weibull function given by (\ref{eq3}).

\vspace{-0.4cm}

\begin{equation}
\label{eq3}
F(t_{stress}) = 1-exp\left[-\left(\frac{t_{stress}}{\eta}\right)^\beta\right]
\end{equation}

where $F(t_{stress})$ is the cumulative probability of breakdowns, $t_{stress}$ is the stress time, $\eta$ is the scale parameter (often referred to as the characteristic lifetime or the time at which 63.2$\%$ of the sample population fails), and $\beta$ is the shape parameter. The Weibull plots for the time to breakdown for the three different dielectrics are shown in Fig. \ref{fig7}(a)-(c). For all three cases, the breakdown is found to be intrinsic, as indicated by $\beta > 1$ for all dielectrics. However, the 600$\degree$C and hybrid dielectrics exhibit much tighter distributions ($\beta > 4$) compared to the 800$\degree$C dielectric, suggesting higher oxide uniformity.

We define the lifetime of the in-situ Al$_2$O$_3$ films as the time to breakdown when the cumulative probability of breakdowns reaches 63$\%$, satisfying $\ln[-\ln(1-F(t_{stress}))]=0$.\cite{9463426} Assuming $E_{BR} \approx V_{BR}/t_{ox}$, the extracted $t_{63\%}$ as a function of $E_{BR}$ for the three different dielectrics is plotted in Fig. \ref{fig7}(d). Lifetime predictions under typical operating conditions are made based on the commonly used E-model \cite{mcpherson1998underlying}, which states that the breakdown time ($t_{BR}$) in the log scale is inversely proportional to the oxide electric fields ($\ln(t_{BR}) \propto -\gamma E_{BR}$). The slope parameter $\gamma$ is often referred to as the electric field acceleration factor, describing the rate at which $t_{BR}$ changes with $E_{BR}$.

For sample A, the estimated time to breakdown ($t_{BR}$) is 10 years if the oxide electric field ($E_{ox}$) remains below 1.5 MV/cm ($E_{ox} \leq 1.5$ MV/cm). Although the lifetime of the 600 $\degree$C dielectric is shorter than that of the 800 $\degree$C dielectric (sample A) at the same stressing electric field, the acceleration factor ($\gamma$) for the 600 $\degree$C dielectric is higher, as evident in Fig. 7(d). This higher acceleration factor is expected due to the higher $\beta$ value of the 600 $\degree$C dielectric. On the other hand, sample B exhibits a $t_{BR}$ of 10 years, with a stressing electric field of 3.5 MV/cm. The longer $t_{BR}$ for the hybrid dielectrics results from reduced polycrystalline features within the bulk of the dielectric compared to thesample A, as well as a pristine dielectric/semiconductor interface compared to the 600 $\degree$C dielectric.

The in-situ Al$_2$O$_3$ grown at 800 $\degree$C (sample A) has fewer near-interface trap states and a higher breakdown field (10 MV/cm) compared to the 600 $\degree$C dielectric (5.6 MV/cm). However, the 600 $\degree$C dielectric shows more uniform time-to-breakdown. Sample B, which combines the advantages of both materials, offers reduced interface traps, higher breakdown strength compared to 600 $\degree$C dielectric, and better uniformity and time-to-breakdown than the 800 $\degree$C dielectrics. Hence, Sample B is a more suitable gate dielectric for $\beta$-Ga$_2$O$_3$-based high-voltage transistor applications. 

In summary, we present in-situ Al$_2$O$_3$ growth via MOCVD at higher temperatures as a superior alternative to conventional deposition methods. We also introduce a hybrid dielectric method, combining an 800 $\degree$C interfacial layer with a 600 $\degree$C bulk dielectric. Capacitance-voltage measurements were used to assess interface quality, revealing that the 800 $\degree$C dielectric outperforms others in near-interface trap states and breakdown field strength. However, the hybrid dielectric (Sample B) shows better uniformity than 800 $\degree$C dielectric and better interface quality than the 600 $\degree$C dielectric, making it a more balanced option. This in-situ dielectric deposition on $\beta$-Ga$_2$O$_3$ promises enhanced performance for future high-performance MOSFETs.

\begin{acknowledgments}
The authors acknowledge the funding from the Coherent/II-VI Foundation Block Gift Program and from the Air Force Office of Scientific Research under award number FA9550-21-0078. Any opinions, findings, conclusions, or recommendations expressed in this article are those of the author(s) and do not necessarily reflect the views of the United States Air Force. A portion of this work was performed in the UCSB Nanofabrication Facility, an open-access laboratory. A portion of this work was performed at the Utah Nanofab cleanroom sponsored by the John and Marcia Price College of Engineering and the Office of the Vice President for Research. We acknowledge discussions with Dr. Ahmad Islam from AFRL Sensors Directorate.
\end{acknowledgments}

\section*{REFERENCES}
\nocite{*}
\bibliography{aipsamp}

\end{document}